This is the author's accepted manuscript of the article published in IEEE Access.

*Digital Object Identifier 10.1109/ACCESS.2025.3633086*

# Comparative Analysis of the Code Generated by Popular Large Language Models (LLMs) for MISRA C++ Compliance

**Malik Muhammad Umer**
College of Computing, Georgia Institute of Technology, Atlanta, GA 30332 USA
e-mail: malikumerpak@gmail.com

**ABSTRACT** Safety-critical systems are engineered systems whose failure or malfunction could result in catastrophic consequences. The software development for safety-critical systems necessitates rigorous engineering practices and adherence to certification standards like DO-178C for avionics. DO-178C is a guidance document which requires compliance to well-defined software coding standards like MISRA C++ to enforce coding guidelines that prevent the use of ambiguous, unsafe, or undefined constructs. Large Language Models (LLMs) have demonstrated significant capabilities in automatic code generation across a wide range of programming languages, including C++. Despite their impressive performance, code generated by LLMs in safety-critical domains must be carefully analyzed for conformance to MISRA C++ coding standards. In this paper, I have conducted a comparative analysis of the C++ code generated by popular LLMs including: OpenAI ChatGPT, Google Gemini, DeepSeek, Meta AI, and Microsoft Copilot for compliance with MISRA C++. The study revealed that none of the evaluated LLMs generated MISRA-compliant code despite clear prompts, with DeepSeek showing the fewest violations and Meta AI the most. While all models could correct individual violations when explicitly instructed, only ChatGPT consistently identified and resolved all targeted rule violations across complete code snippets, whereas others achieved partial success. Overall, LLMs show promise as aids for initial code generation, but they are not yet dependable for producing fully MISRA-compliant code required in safety-critical domains.

## I. INTRODUCTION

Safety-critical systems are engineered systems whose failure or malfunction could result in catastrophic consequences, including loss of human life, significant environmental damage, or major financial loss [1]. These systems are commonly found in domains such as aerospace, automotive, medical devices, railways, and nuclear energy. The integrity, reliability, and predictable behavior of safety-critical systems are paramount, as even minor defects can lead to severe outcomes. As these systems increasingly rely on complex software components, ensuring their safety becomes both a technical and regulatory challenge.

The software development for safety-critical systems necessitates rigorous engineering practices applied to the design, implementation, verification, and validation of the software. This requires high assurance techniques, such as formal methods, thorough testing, traceability, and adherence to stringent standards like DO-178C for avionics, ISO 26262 for automotive [20], and IEC 62304 for medical devices [21]. The goal is to minimize software-induced hazards and ensure that the software behaves correctly under all foreseeable conditions. As the complexity of embedded software grows, so too does the importance of disciplined processes and safety assurance frameworks in its development lifecycle.

DO-178C [2], formally titled Software Considerations in Airborne Systems and Equipment Certification, is a guidance document developed by RTCA, with its European counterpart ED-12C [22] published by EUROCAE, to ensure the safety and reliability of software used in airborne systems. It is widely recognized and accepted by aviation regulatory authorities, such as the FAA and EASA, for certifying software in civil aircraft. DO-178C defines a structured and rigorous framework for the software development life cycle, emphasizing traceability, verification, and risk-based assurance activities. While DO-178C does not mandate specific programming languages or coding styles, it requires adherence to well-defined and verifiable software coding standards, such as MISRA C++ and CERT C++, to prevent the use of ambiguous, unsafe, or undefined constructs.

MISRA C++ [3] document developed by Motor Industry Software Reliability Association (MISRA) specifies a set of guidelines for the use of C++ language in critical systems. By following MISRA C++, developers can ensure that their code adheres to established best practices, mitigating the risk of errors and vulnerabilities in software. The specific revision used for this study is MISRA C++:2008, published in June 2008, which explains the total set of 226 rules and recommendations distributed in 70 categories.

Every MISRA C++ rule is classified as "Required", "Advisory" or "Document". "Required" rules are mandatory requirements placed on the developer. C++ code that is claimed to conform to MISRA C++ shall comply with every "Required" rule. "Advisory" rules are requirements placed on the developer that should normally be followed. However, they do not have the mandatory status of "Required" rules. "Document" rules are mandatory requirements placed on the developer whenever the associated feature is used within code. Deviations are not permitted against this class of rule. MISRA C++:2008 contains 198 "Required" rules, 18 "Advisory" rules, and 12 "Document" rules.

Conformance to coding standards such as MISRA C++ is evaluated using Static Analysis Tools (SAT), which systematically examine the source code without executing it to ensure adherence to the prescribed guidelines. By scanning and analyzing the source code, SAT can identify violations of the prescribed coding standards or guidelines, so that they can be fixed to make the source code clean.

Large Language Models (LLMs), such as OpenAI ChatGPT [4], DeepSeek [5], Google Gemini [6], Meta AI [7] and Microsoft Copilot [8], have demonstrated significant capabilities in automatic code generation by leveraging extensive training on vast corpora of natural language and programming language data. These models utilize deep learning architectures, primarily based on transformer networks, to learn contextual patterns and semantic structures inherent in source code. As a result, LLMs can generate syntactically correct and semantically meaningful code snippets across a wide range of programming languages, including Python, C++, Java, and JavaScript. The ability of LLMs to understand and synthesize code based on natural language prompts allows developers to express intent in plain language and receive executable source code in response.

Despite their impressive performance, the code generated by LLMs must be carefully analyzed for compliance with coding standards including MISRA C++ especially in safety-critical domains, since DO-178C specifies that relevant coding standards must be defined and followed for certification of avionics software [2].

To explore the problem stated above, I have analyzed how compliant is the code generated by popular LLMs with MISRA C++. Furthermore, I have compared the results of MISRA C++ violations produced by the analysis of the code using Static Analysis Tool for each LLM. 5 popular LLMs including OpenAI ChatGPT, DeepSeek, Google Gemini, Meta AI, and Microsoft Copilot were used to generate code for calculation of checksum, which is commonly used functionality in safety-critical applications for validation of critical data transferred across communication buses. A popular Static Analysis Tool (SAT), PC-lint Plus [9], was used to analyze the source code generated by LLMs for compliance with MISRA C++ coding standard.

The remainder of this paper is organized into following sections. Section II describes the methodology and framework adopted to analyze and compare the MISRA C++ compliance of code generated by five widely used LLMs. Section III presents and analyzes the results of the comparative evaluation. Section IV reviews related work in this domain. Section V outlines the main threats to validity identified during the study and highlights potential areas that warrant further exploration, implementation, and assessment. Finally, Section VI summarizes the key findings and conclusions of the study.

## II. COMPARATIVE ANALYSIS

This section outlines the methodology and framework employed in this study to analyze and compare the MISRA C++ compliance of code generated by 5 widely used LLMs.

### A. LLMs AND GENERATED CODE

For this study, I have selected 5 popular LLMs with free and online (web-based) versions including: OpenAI ChatGPT GPT-4o [4], DeepSeek-V3 [5], Google Gemini 2.0 Flash [6], Meta AI [7] and Microsoft Copilot [8].

**TABLE 1.** **Summary of code generated by LLMs.**

| LLM | Lines of Code | Algorithms | Build Status |
|---|---|---|---|
| ChatGPT | 69 | Internet, Fletcher-16, and CRC-32 | Successful |
| DeepSeek | 60 | CRC-32, Adler-32, and Fletcher-16 | Failed (2 errors) |
| Gemini | 65 | Mod Sum, Fletcher-16, and CRC-8 | Successful |
| Meta AI | 52 | CRC-8, CRC-16, and Adler-32 | Successful (3 warnings) |
| Copilot | 68 | XOR and CRC-32 | Successful (2 warnings) |

All LLMs were given the following prompt in natural language to generate MISRA C++ compliant code for popular checksum algorithms: "Generate a C++ code which implements popular Checksum algorithms. The code should be 100% compliant with MISRA C++ standard".

**TABLE 2. Summary of violations of MISRA C++ rules for the source code generated by LLMs including OpenAI ChatGPT, DeepSeek, Google Gemini, Meta AI, and Microsoft Copilot based on analysis by PC-lint Plus.**

| Rule | Category | Description | ChatGPT | DeepSeek | Gemini | Meta AI | Copilot |
|---|---|---|---|---|---|---|---|
| Rule 0–1–3 | Required | A project shall not contain unused variables. | - | - | - | 3 | 2 |
| Rule 0–1–4 | Required | A project shall not contain non-volatile Plain Old Data (POD) variables having only one use. | - | - | - | 3 | 2 |
| Rule 0–1–6 | Required | A project shall not contain instances of non-volatile variables being given values that are never subsequently used. | - | - | - | 3 | - |
| Rule 0–1–9 | Required | There shall be no dead code. | - | - | - | 3 | - |
| Rule 3–3–1 | Required | Objects or functions with external linkage shall be declared in a header file. | - | - | - | 3 | 3 |
| Rule 3–9–2 | Advisory | typedefs that indicate size and signedness should be used in place of the basic numerical types. | 1 | - | 2 | - | - |
| Rule 5–0–2 | Advisory | Limited dependence should be placed on C++ operator precedence rules in expressions. | 1 | - | - | - | - |
| Rule 5–0–3 | Required | A cvalue expression shall not be implicitly converted to a different underlying type. | 1 | - | - | - | 2 |
| Rule 5–0–4 | Required | An implicit integral conversion shall not change the signedness of the underlying type. | 8 | 5 | 10 | 21 | - |
| Rule 5–0–6 | Required | An implicit integral or floating-point conversion shall not reduce the size of the underlying type. | 4 | - | 2 | 4 | - |
| Rule 5–0–9 | Required | An explicit integral conversion shall not change the signedness of the underlying type of a cvalue expression. | 1 | - | - | - | - |
| Rule 5–0–10 | Required | If the bitwise operators ~ and << are applied to an operand with an underlying type of unsigned char or unsigned short, the result shall be immediately cast to the underlying type of the operand. | 1 | 1 | 2 | 2 | 2 |
| Rule 5–0–13 | Required | The condition of an if-statement and the condition of an iteration statement shall have type bool. | - | - | 1 | 2 | - |
| Rule 5–0–20 | Required | Non-constant operands to a binary bitwise operator shall have the same underlying type. | 1 | 1 | - | 1 | 2 |
| Rule 5–0–21 | Required | Bitwise operators shall only be applied to operands of unsigned underlying type. | - | - | 6 | 9 | - |
| Rule 5–2–12 | Required | An identifier with array type passed as a function argument shall not decay to a pointer. | - | - | - | 3 | 2 |
| Rule 6–6–5 | Required | A function shall have a single point of exit at the end of the function. | - | - | 1 | - | - |
| Rule 7–1–1 | Required | A variable which is not modified shall be const qualified. | - | 6 | 7 | 7 | 7 |
| Rule 7–3–1 | Required | The global namespace shall only contain main, namespace declarations and extern "C" declarations. | - | - | - | 3 | 3 |
| Rule 16–0–1 | Required | #include directives in a file shall only be preceded by other pre-processor directives or comments. | - | - | 1 | - | - |
| | | **Total:** | 18 | 13 | 32 | 67 | 25 |

Checksum computation routines are commonly implemented functionalities in safety-critical applications such as avionics, automotive, and industrial control systems. They serve a vital function in verifying the integrity and accuracy of critical data transmitted across communication buses or between embedded subsystems. Such routines are integral to system safety mechanisms required by standards like DO-178C, ISO 26262, and IEC 61508, which emphasize deterministic and verifiable software behavior for data validation and transmission integrity. Moreover, checksum algorithms are public-domain and platform-independent, allowing transparent replication of results using static analysis tools (e.g., PC-lint Plus) without encountering proprietary or export-restricted code, a frequent limitation in safety-critical research. Their structure and purpose closely reflect real-world implementations used in certified aerospace and automotive systems, ensuring that the analysis presented in this study remains representative of practical industrial safety practices.

The code generated by ChatGPT, DeepSeek, Gemini, Meta AI, and Copilot respectively can be accessed from GitHub [10]. TABLE 1. presents a summary of the code generated by all LLMs, including the number of lines of code, the implemented checksum algorithms, and the corresponding build status.

### *B. SAT AND MISRA C++ COMPLIANCE*

PC-lint Plus SAT [9] was used to analyze the code generated by all LLMs by giving the following terminal command: "pclp64 co-gcc.lnt lnt\au-misra-cpp.lnt -std=c++11 -os(<output_file_path>) -summary <source_file_path>". The list of MISRA C++:2008 violations for the code generated by ChatGPT, DeepSeek, Gemini, Meta AI, and Copilot respectively can be accessed from GitHub [11].

## III. RESULTS AND DISCUSSION

This section presents and analyzes the results of the MISRA C++ rule violations comparison for safety-critical code generated by ChatGPT, DeepSeek, Gemini, Meta AI, and Copilot. Initially, a summary of violations reported by PC-lint Plus is provided, followed by a detailed analysis and comparative evaluation of the findings.

### *A. SUMMARY OF RESULTS*

The summary of MISRA C++ rule violations for C++ code generated by OpenAI ChatGPT, DeepSeek, Google Gemini, Meta AI and Microsoft Copilot is presented in TABLE 2. The results demonstrate that none of the evaluated large language models (LLMs) generate code that is fully compliant with MISRA C++ guidelines, even when explicit instructions to ensure compliance are included in the prompt.

Consequently, the code produced by these models cannot be deemed reliable for use in safety-critical applications without thorough manual review or MISRA C++ compliance check using a static analysis tool (SAT) prior to deployment. Among the evaluated models, the code generated by DeepSeek exhibited the lowest number of MISRA C++ rule violations, whereas the code produced by Meta AI demonstrated the highest number of violations. Specifically, the number of violations identified for ChatGPT, DeepSeek, Gemini, Meta AI, and Copilot were 18, 13, 32, 67, and 25, respectively. FIGURE 1. presents a comparative overview of the total number of MISRA C++ rule violations associated with the source code generated by each LLM.

Furthermore, the source code generated by DeepSeek exhibited violations in the fewest number of distinct MISRA C++ rules, whereas the code produced by Meta AI showed violations across the highest number of distinct rules. Specifically, the number of different MISRA C++ rules for which at least one violation was identified in the code generated by ChatGPT, DeepSeek, Gemini, Meta AI, and Copilot were 8, 4, 9, 14, and 9, respectively. FIGURE 2. provides a comparative analysis of the total number of distinct MISRA C++ rules violated in the source code generated by each LLM.

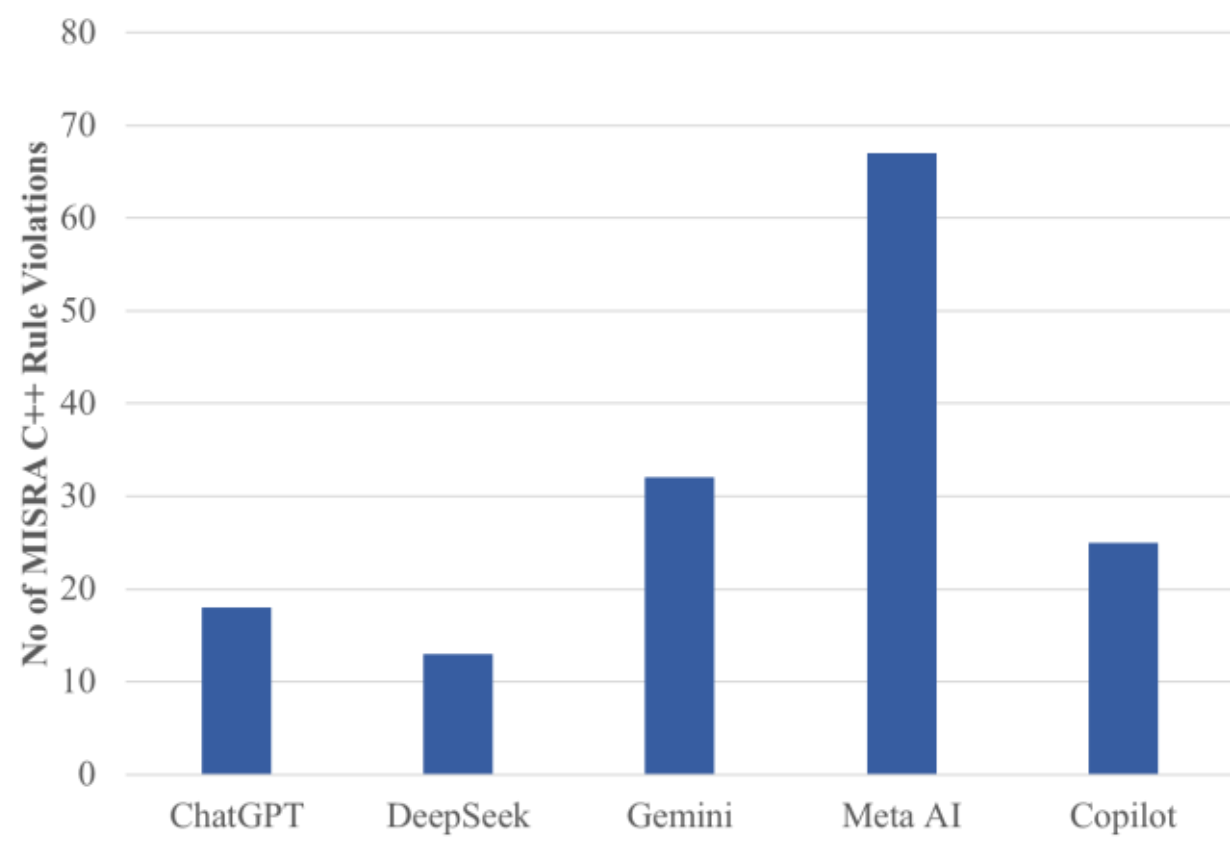

**FIGURE 1. Total No of violations of MISRA C++ rules for the source code generated by each LLM.**

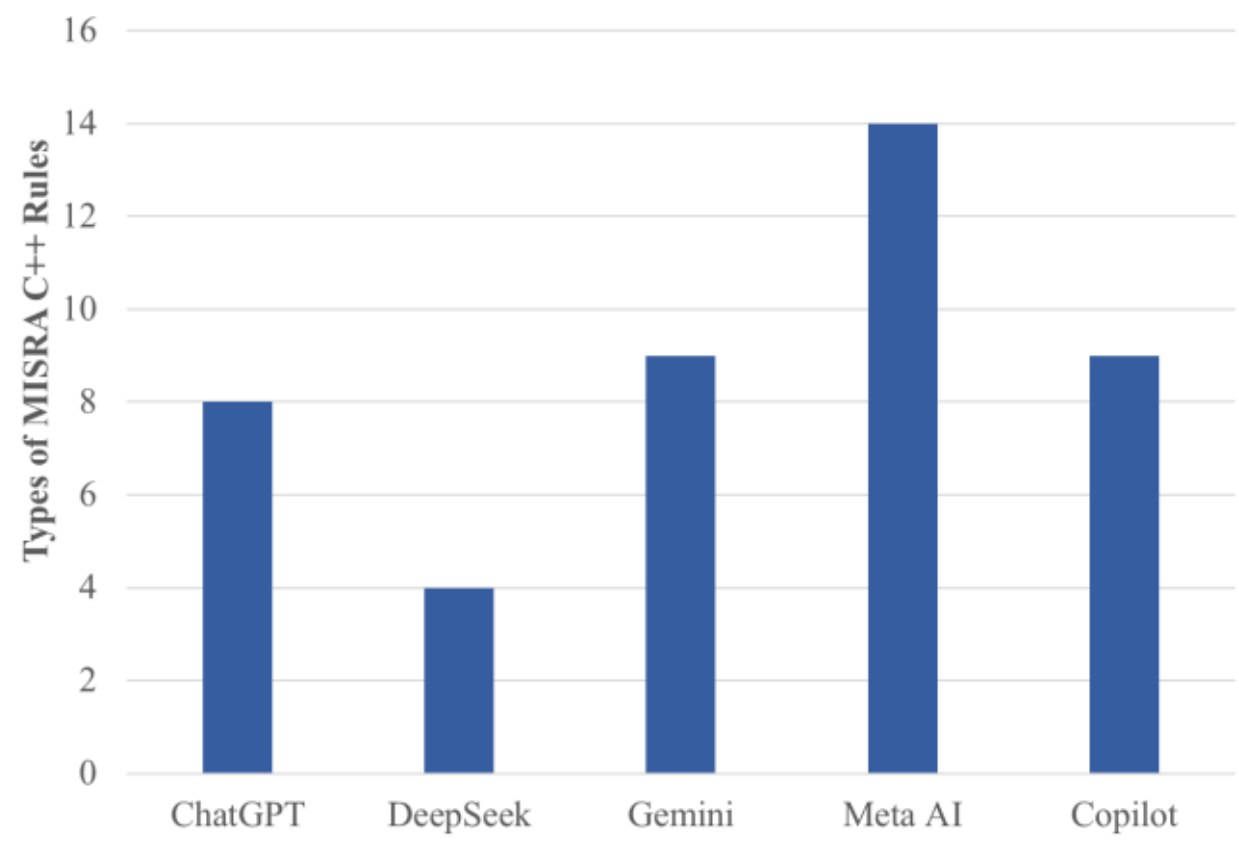

**FIGURE 2. Total count of different types of MISRA C++ rules violated for the source code generated by each LLM.**

The highest number of violations were recorded for the Rule 5-0-4, with a total of 44 instances, followed by 27

violations of Rule 7-1-1 and 15 violations of Rule 5-0-21. Rule 5-0-4 relates to undesirable implicit signed to unsigned type conversions, which may lead to undefined behavior. Rule 7-1-1 requires that a variable which does not need to be modified, shall be declared with const qualification so that it cannot be modified anywhere in the code. Rule 5-0-21 highlights that the Bitwise operations (~, <<, <<=, >>, >>=, &, &=, ^, ^=, | and |=) are not meaningful on signed underlying type.

In contrast, the lowest number of violations, one each, was observed for Rules 5-0-2, 5-0-9, 6-6-5, and 16-0-1. Summary of MISRA C++:2008 rules and their corresponding number of violations is provided in TABLE 3. In total, 20 distinct rules were violated out of 226 defined by the standard, comprising 18 "Required" rules and 2 "Advisory" rules.

**TABLE 3.** Total No of violations for each MISRA C++ rule.

| MISRA C++ Rule | Total No of Violations |
|---|---|
| Rule 0–1–3 | 5 |
| Rule 0–1–4 | 5 |
| Rule 0–1–6 | 3 |
| Rule 0–1–9 | 3 |
| Rule 3–3–1 | 6 |
| Rule 3–9–2 | 3 |
| Rule 5–0–2 | 1 |
| Rule 5–0–3 | 3 |
| Rule 5–0–4 | 44 |
| Rule 5–0–6 | 10 |
| Rule 5–0–9 | 1 |
| Rule 5–0–10 | 8 |
| Rule 5–0–13 | 3 |
| Rule 5–0–20 | 5 |
| Rule 5–0–21 | 15 |
| Rule 5–2–12 | 5 |
| Rule 6–6–5 | 1 |
| Rule 7–1–1 | 27 |
| Rule 7–3–1 | 6 |
| Rule 16–0–1 | 1 |

### B. ANALYSIS OF CHATGPT

The source code generated by ChatGPT consists of 69 lines and includes violations of 18 MISRA C++:2008 rules. Specifically, it contains 8 violations of Rule 5-0-4, 4 violations of Rule 5-0-6, and 1 violation each of Rules 3-9-2, 5-0-2, 5-0-3, 5-0-9, 5-0-10, and 5-0-20.

ChatGPT generates no violation of Rule 7-1-1 contrary to multiple violations of the same by all other LLMs, which requires that a variable which does not need to be modified, shall be declared with const qualification so that it cannot be modified.

A detailed breakdown of these violations is presented in TABLE 4. with the corresponding code highlighted using red font and underlined to indicate specific instances of non-compliance. Notably, the 8 violations of Rule 5-0-4 and the 4 violations of Rule 5-0-6 occur at the same line of code and arise from a common construct, indicating a strong interrelation between the two rules in this context.

**TABLE 4.** Details of MISRA C++:2008 rule violations for the source code generated by OpenAI ChatGPT.

| Rule | Total | Violation |
|---|---|---|
| Rule 5–0–4 | 8 | **Problem:** Implicit conversion from 'short' to 'unsigned char' changes signedness of underlying type.<br>**Code:** constexpr std::array <std::uint8_t, 8U> testData = {0xDE, 0xAD, 0xBE, 0xEF, 0x12, 0x34, 0x56, 0x78}; |
| Rule 5–0–6 | 4 | **Problem:** Implicit conversion of integer to smaller underlying type ('short' to 'unsigned char').<br>**Code:** constexpr std::array <std::uint8_t, 8U> testData = {0xDE, 0xAD, 0xBE, 0xEF, 0x12, 0x34, 0x56, 0x78}; |
| Rule 3–9–2 | 1 | **Problem:** Use of modifier or type 'int' outside of a typedef.<br>**Code:** return static_cast<int> (checksum1 ^ checksum2 ^ checksum3); |
| Rule 5–0–2 | 1 | **Problem:** Dependence placed on C++ operator precedence.<br>**Code:** if (i + 1U < data.size()) |
| Rule 5–0–3 | 1 | **Problem:** Implicit conversion of underlying type of integer cvalue expression from 'unsigned short' to 'unsigned int'.<br>**Code:** return static_cast<int> (checksum1 ^ checksum2 ^ checksum3); |
| Rule 5–0–9 | 1 | **Problem:** Cast of integer cvalue expression changes signedness.<br>**Code:** return static_cast<int> (checksum1 ^ checksum2 ^ checksum3); |
| Rule 5–0–10 | 1 | **Problem:** The result of the << operator applied to an object with an underlying type of 'unsigned short' must be cast to 'unsigned short' in this context.<br>**Code:** return static_cast<std::uint16_t> ((sum2 << 8U) \| sum1); |
| Rule 5–0–20 | 1 | **Problem:** Bitwise operator '^' used with non-constant operands of differing underlying types ('unsigned short' and 'unsigned int').<br>**Code:** return static_cast<int> (checksum1 ^ checksum2 ^ checksum3); |

The violations for Rules 5-0-4 and 5-0-6 were identified in a test example code and flagged by PC-lint Plus with messages: "implicit conversion from 'short' to 'unsigned char' changes signedness of underlying type" (for Rule 5-0-4) and "implicit conversion of integer to smaller underlying type ('short' to 'unsigned char')" (for Rule 5-0-6). These messages pertain to the line "constexpr std::array<std::uint8_t, 8U> testData = {0xDE, 0xAD, 0xBE, 0xEF, 0x12, 0x34, 0x56, 0x78};" where each array value triggers the flagged conversion.

There is one violation each of Rules 3-9-2 and 5-0-2, both of which are classified as "Advisory" rules under MISRA

C++:2008. As adherence to advisory rules is not mandatory, these violations may be considered non-critical and potentially disregarded. Furthermore, a single instance of non-compliance with Rules 5-0-3, 5-0-9, and 5-0-20 was identified on the same line of test code. The violation occurs in the reference code: “return static_cast<int> (checksum1 ^ checksum2 ^ checksum3);” within “main()” function, which is used to test the checksum algorithms.

The remaining single violation pertains to MISRA C++:2008 Rule 5-0-10 and is associated with the core implementation of the checksum algorithms. Specifically, the infraction occurs on the statement: “return static_cast <std::uint16_t> ((sum2 << 8U) | sum1);”. This non-compliance arises due to the application of the left-shift operator to an object of type “std::uint16_t” without explicitly casting the result back to the same type, as mandated by the rule to prevent unintended type promotion.

**INPUT PROMPT**

The following line of code:

```
return static_cast<std::uint16_t>((sum2 << 8U) | sum1);
```

violates MISRA C++:2008 Rule 5-0-10, which states:
"The result of the << operator applied to an object with an underlying type of 'unsigned short' must be cast to 'unsigned short' in this context.“

Please revise the code below to ensure compliance with Rule 5-0-10:

```
// Fletcher-16 Checksum
   std::uint16_t fletcher16(const std::array<std::uint8_t, 8U>& data)
   {
      std::uint16_t sum1 = 0U;
      std::uint16_t sum2 = 0U;

      for (std::size_t i = 0U; i < data.size(); ++i)
      {
         sum1 = static_cast<std::uint16_t>((sum1 + data[i]) % 255U);
         sum2 = static_cast<std::uint16_t>((sum2 + sum1) % 255U);
      }
      return static_cast<std::uint16_t>((sum2 << 8U) | sum1);
   }
```

**OUTPUT RESPONSE**

To ensure compliance with MISRA C++:2008 Rule 5-0-10, you need to explicitly cast the result of the shift operation (sum2 << 8U) back to std::uint16_t before performing the bitwise OR. This prevents unintended type promotion during the shift.

Here is the revised code with the required fix:

```
// Fletcher-16 Checksum
   std::uint16_t fletcher16(const std::array<std::uint8_t, 8U>& data)
   {
      std::uint16_t sum1 = 0U;
      std::uint16_t sum2 = 0U;

      for (std::size_t i = 0U; i < data.size(); ++i)
      {
         sum1 = static_cast<std::uint16_t>((sum1 + data[i]) % 255U);
         sum2 = static_cast<std::uint16_t>((sum2 + sum1) % 255U);
      }
      return static_cast<std::uint16_t>(
         static_cast<std::uint16_t>(sum2 << 8U) | sum1
         );
   }
```

☑ Explanation:

- sum2 << 8U promotes sum2 (a std::uint16_t) to int during the shift.
- MISRA Rule 5-0-10 requires explicitly casting this result back to std::uint16_t before using it in further expressions.
- The fix is done by adding static_cast<std::uint16_t>(sum2 << 8U).

This modification ensures the code adheres to the required MISRA rule and avoids implicit type promotions.

FIGURE 3. **Input prompt and output response of ChatGPT to generate the code with MISRA C++:2008 rule 5-0-10 fixed.**

To further evaluate ChatGPT’s effectiveness in addressing MISRA C++:2008 non-compliance, a targeted prompt was issued to correct the Rule 5-0-10 violation. In response, ChatGPT produced a revised version of code that resolved the identified issue, as illustrated in FIGURE 3.

### *C. ANALYSIS OF DEEPSEEK*

The source code generated by DeepSeek comprises 60 lines and exhibits 13 instances of non-compliance with MISRA C++:2008 guidelines. Specifically, it includes 5 violations of Rule 5-0-4, 6 violations of Rule 7-1-1, and 1 violation each of Rules 5-0-10 and 5-0-20.

Among all evaluated LLMs, DeepSeek generates minimum number of MISRA C++ rule violations (13) and for a minimum number of distinct MISRA C++ rules (4). A detailed breakdown of these violations is presented in TABLE 5. with the corresponding code highlighted using red font and underlined to indicate the specific instances of non-compliance. It is noteworthy that all five violations of Rule 5-0-4 originate from a single line of code.

TABLE 5. **Details of MISRA C++:2008 rule violations for the source code generated by DEEPSEEK.**

| Rule | Total | Violation |
|---|---|---|
| Rule 5–0–4 | 5 | **Problem:** Implicit conversion from 'signed char' to 'unsigned char' changes signedness of underlying type.<br>**Code:** const std::array<uint8_t, 5> testData = { 'H', 'e', 'l', 'l', 'o' }; |
| Rule 7–1–1 | 6 | **Problem:** Parameter 'data' of function 'Checksum::crc32 (const uint8_t *, uint32_t)' could be const.<br>**Code:** uint32_t crc32(const uint8_t* data, uint32_t length) noexcept { |
| | | **Problem:** Parameter 'length' of function 'Checksum::crc32 (const uint8_t *, uint32_t)' could be const.<br>**Code:** uint32_t crc32(const uint8_t* data, uint32_t length) noexcept { |
| | | **Problem:** Parameter 'data' of function 'Checksum::adler32 (const uint8_t *, uint32_t)' could be const.<br>**Code:** uint32_t adler32 (const uint8_t* data, uint32_t length) noexcept { |
| | | **Problem:** Parameter 'length' of function 'Checksum::adler32 (const uint8_t *, uint32_t)' could be const.<br>**Code:** uint32_t adler32 (const uint8_t* data, uint32_t length) noexcept { |
| | | **Problem:** Parameter 'data' of function 'Checksum::fletcher16 (const uint8_t *, uint32_t)' could be const.<br>**Code:** uint16_t fletcher16 (const uint8_t* data, uint32_t length) noexcept { |
| | | **Problem:** Parameter 'length' of function 'Checksum::fletcher16 (const uint8_t *, uint32_t)' could be const.<br>**Code:** uint16_t fletcher16 (const uint8_t* data, uint32_t length) noexcept { |
| Rule 5–0–10 | 1 | **Problem:** The result of the << operator |
| | | applied to an object with an underlying type of 'unsigned short' must be cast to 'unsigned short' in this context.<br>**Code:** return (sum2 << 8U) \| sum1; |
| Rule 5–0–20 | 1 | **Problem:** Bitwise operator '^' used with non-constant operands of differing underlying types ('unsigned int' and 'unsigned char').<br>**Code:** crc = (crc >> 8U) ^ CRC32_TABLE [(crc ^ data[i]) & 0xFFU]; |

The Rule 5-0-4 violations were detected in test example code and were flagged by PC-lint Plus with the diagnostic message: "implicit conversion from 'signed char' to 'unsigned char' changes signedness of underlying type". This pertains to the line "const std::array<uint8_t, 5> testData = { 'H', 'e', 'l', 'l', 'o' };", where each character triggers conversion issue.

The remaining 6 violations of Rule 7-1-1, along with 1 violation each of Rules 5-0-10 and 5-0-20, were identified within the core implementation of the checksum algorithms. To further assess DeepSeek's capability in rectifying MISRA C++:2008 non-compliance, a focused prompt was issued to address the Rule 7-1-1 infraction. In response, DeepSeek generated a corrected version of the code, which successfully resolved the non-compliance, as presented in FIGURE 4.

**INPUT PROMPT**

The following line of code:
uint16_t fletcher16(const uint8_t* data, uint32_t length) noexcept {

violates MISRA C++:2008 Rule 7-1-1, which states:
"parameter 'length' of function 'Checksum::fletcher16(const uint8_t *, uint32_t)' could be const ".

Please revise the code below to ensure compliance with Rule 7-1-1:

```
// Fletcher-16 implementation
  uint16_t fletcher16(const uint8_t* data, uint32_t length) noexcept {
    uint16_t sum1 = 0U;
    uint16_t sum2 = 0U;
    for (uint32_t i = 0U; i < length; ++i) {
      sum1 = (sum1 + data[i]) % 255U;
      sum2 = (sum2 + sum1) % 255U;
    }
    return (sum2 << 8U) | sum1;
  }
```

**OUTPUT RESPONSE**

Here's the revised code that fixes MISRA C++ Rule 7-1-1 by making the length parameter const since it is not modified inside the function:

Fixed Code (MISRA C++:2008 Rule 7-1-1 Compliance)

```
#include <cstdint>

// Fletcher-16 implementation
uint16_t fletcher16(const uint8_t* data, const uint32_t length) noexcept
{
  uint16_t sum1 = 0U;
  uint16_t sum2 = 0U;

  for (uint32_t i = 0U; i < length; ++i)
  {
    sum1 = (sum1 + data[i]) % 255U;
    sum2 = (sum2 + sum1) % 255U;
  }

  return (sum2 << 8U) | sum1;
}
```

Key Changes:

1. Added const to length parameter. Since length is only read and never modified, making it const complies with MISRA Rule 7-1-1. This improves code safety by preventing accidental modifications.

2. Kept the rest of the logic unchanged. The function remains efficient and correct while now following MISRA guidelines.

**FIGURE 4.** Input prompt and output response of DeepSeek to generate the code with MISRA C++:2008 rule 7-1-1 fixed.

### D. ANALYSIS OF GEMINI

The source code generated by Google Gemini consists of 65 lines and includes 32 instances of non-compliance with the MISRA C++:2008 standard. Specifically, the code contains 10 violations of Rule 5-0-4, 7 violations of Rule 7-1-1, 6 violations of Rule 5-0-21, 2 violations each of the Rules 3-9-2, 5-0-6, and 5-0-10, and 1 violation each of Rules 5-0-13, 6-6-5, and 16-0-1. A comprehensive summary of these violations is provided in TABLE 6. where the relevant portions of the code are underlined and displayed in red font to clearly indicate the specific instances of non-conformance.

Violation of Rule 16-0-1 pertains to the implementation in the "main()" function of test code. This violation arises due to its inclusion in a shared scope; it would not occur if "main()" function were placed in an isolated scope. The 2 violations of Rule 3-9-2 are categorized as "Advisory" under MISRA C++:2008, indicating that adherence to this rule is recommended but not mandatory. Of the 10 violations recorded for Rule 5-0-4, 2 occur on the same line: "for (std::uint8_t i = 0; i < 8; ++i)". Similarly, both violations of Rule 5-0-21 appear on the line: "crc = (crc << 1) ^ 0x07".

**TABLE 6.** Details of MISRA C++:2008 rule violations for the source code generated by Google Gemini.

| Rule | Total | Violation |
|---|---|---|
| Rule 5–0–4 | 10 | **Problem:** Implicit conversion from 'signed char' to 'unsigned char' changes signedness of underlying type.<br>**Code:** crc = (crc << 1) ^ 0x07; |
| | | **Problem:** Implicit conversion from 'short' to 'unsigned short' changes signedness of underlying type.<br>**Code:** return static_cast<std::uint8_t>(sum & 0xFF); |
| | | **Problem:** Implicit conversion from 'signed char' to 'std::uint16_t' (aka 'unsigned short') changes signedness of underlying type.<br>**Code:** std::uint16_t sum1 = 0; |
| | | **Problem:** Implicit conversion from 'signed char' to 'std::uint16_t' (aka 'unsigned short') changes signedness of underlying type.<br>**Code:** std::uint16_t sum2 = 0; |
| | | **Problem:** Implicit conversion from 'short' to 'unsigned short' changes signedness of underlying type.<br>**Code:** sum1 = (sum1 + byte) % 255; |
| | | **Problem:** Implicit conversion from 'short' to 'unsigned short' changes signedness of underlying type.<br>**Code:** sum2 = (sum2 + sum1) % 255; |
| | | **Problem:** Implicit conversion from 'short' to 'std::uint8_t' (aka 'unsigned char') changes signedness of underlying type.<br>**Code:** std::uint8_t crc = 0xFF; |
| | | **Problem:** Implicit conversion from 'signed char' to 'std::uint8_t' (aka 'unsigned char') changes signedness of underlying type.<br>**Code:** for (std::uint8_t i = 0; i < 8; ++i) { |
| | | **Problem:** Implicit conversion from 'signed |
| | | char' to 'unsigned char' changes signedness of underlying type.<br>**Code:** for (std::uint8_t i = 0; i < 8; ++i) { |
| | | **Problem:** Implicit conversion from 'unsigned char' to 'short' changes signedness of underlying type.<br>**Code:** if (crc & 0x80) { |
| Rule 7–1–1 | 7 | **Problem:** Local variable 'my_data' could be const.<br>**Code:** std::vector<std::uint8_t> my_data = {0x01, 0x02, 0x03, 0x04}; |
| | | **Problem:** Local variable 'simple_checksum' could be const.<br>**Code:** std::uint8_t simple_checksum = Checksum::simpleChecksum(my_data); |
| | | **Problem:** Local variable 'fletcher_checksum' could be const.<br>**Code:** std::uint16_t fletcher_checksum = Checksum::fletcher16(my_data); |
| | | **Problem:** Local variable 'crc_checksum' could be const.<br>**Code:** std::uint8_t crc_checksum = Checksum::crc8 (my_data); |
| | | **Problem:** Local variable 'byte' could be const.<br>**Code:** for (std::uint8_t byte : data) { |
| | | **Problem:** Local variable 'byte' could be const.<br>**Code:** for (std::uint8_t byte : data) { |
| | | **Problem:** Local variable 'sum' could be const.<br>**Code:** std::uint16_t sum = std::accumulate (data.begin(), data.end(), static_cast <std::uint16_t>(0)); |
| Rule 5–0–21 | 6 | **Problem:** Bitwise operator '^' applied to signed underlying type.<br>**Code:** crc = (crc << 1) ^ 0x07; |
| | | **Problem:** Bitwise operator '<<' applied to signed underlying type.<br>**Code:** crc = (crc << 1) ^ 0x07; |
| | | **Problem:** Bitwise operator '<<=' applied to signed underlying type.<br>**Code:** crc <<= 1; |
| | | **Problem:** Bitwise operator '&' applied to signed underlying type.<br>**Code:** return static_cast<std::uint8_t>(sum & 0xFF); |
| | | **Problem:** Bitwise operator '<<' applied to signed underlying type.<br>**Code:** return (sum2 << 8) \| sum1; |
| | | **Problem:** Bitwise operator '&' applied to signed underlying type.<br>**Code:** if (crc & 0x80) { |
| Rule 3–9–2 | 2 | **Problem:** Use of modifier or type 'int' outside of a typedef.<br>**Code:** std::cout << "Simple Checksum: 0x" << std::hex << static_cast<int> (simple_checksum) << std::endl; |
| | | **Problem:** Use of modifier or type 'int' outside of a typedef.<br>**Code:** std::cout << "CRC-8 Checksum: 0x" << std::hex << static_cast<int> (crc_checksum) << std::endl; |
| Rule 5–0–6 | 2 | **Problem:** Implicit conversion of integer to smaller underlying type ('short' to 'std::uint8_t' (aka 'unsigned char')).<br>**Code:** std::uint8_t crc = 0xFF; |
| | | **Problem:** Implicit conversion of integer to smaller underlying type ('short' to 'bool').<br>**Code:** if (crc & 0x80) { |
| Rule 5–0–10 | 2 | **Problem:** The result of the << operator applied to an object with an underlying type of 'unsigned char' must be cast to 'unsigned char' in this context.<br>**Code:** crc = (crc << 1) ^ 0x07; |
| | | **Problem:** The result of the << operator applied to an object with an underlying type of 'unsigned short' must be cast to 'unsigned short' in this context.<br>**Code:** return (sum2 << 8) \| sum1; |
| Rule 5–0–13 | 1 | **Problem:** Condition of 'if' statement has non-Boolean type 'int'.<br>**Code:** if (crc & 0x80) { |
| Rule 6–6–5 | 1 | **Problem:** 'return' statement before end of function 'main'.<br>**Code:** return 1; |
| Rule 16–0–1 | 1 | **Problem:** Declaration of 'Checksum' before #include.<br>**Code:** namespace Checksum { |

Violations of Rules 16-0-1 and 6-6-5 were observed in code generated by Google Gemini, with no such occurrences in code produced by the other LLMs. Moreover, all violations of Rules 3-9-2, 16-0-1, 6-6-5, and the 4 instances of Rule 7-1-1 all relate to the implementation in the "main()" function. Violations of Rules 5-0-13 and 5-0-21 were identified only in the code generated by Gemini and Meta AI. Notably, 8 out of the 10 violations of Rule 5-0-4 in Gemini's output were found in the core implementation, whereas in ChatGPT and DeepSeek, similar violations were limited to the test code.

To further evaluate Gemini's ability to address MISRA C++:2008 non-compliance, a targeted prompt was issued to specifically correct the Rule 6-6-5 violation. In response, Gemini successfully generated a revised version of complete code that resolved the infraction, as illustrated in FIGURE 5.

### *E. ANALYSIS OF META AI*

The source code generated by Meta AI comprises 52 lines and exhibits a total of 67 violations of the MISRA C++:2008 standard. Specifically, it includes 21 violations of Rule 5-0-4, 9 violations of Rule 5-0-21, 7 violations of Rule 7-1-1, 4 violations of Rule 5-0-6, and 3 violations each of Rules 0-1-3, 0-1-4, 0-1-6, 0-1-9, 3-3-1, 5-2-12, and 7-3-1. Additionally, there are 2 violations each of Rules 5-0-10 and 5-0-13, and 1 violation of Rule 5-0-20. Among all evaluated large language models (LLMs), Meta AI produced the highest number of MISRA C++ rule violations (67) as well as the greatest variety, with infractions spanning 14 distinct rules.

A detailed summary of these violations is presented in TABLE 7. where the relevant code segments are underlined and highlighted in red to clearly indicate the specific instances of non-compliance.

**INPUT PROMPT**

The following line of code:
return 1;

violates MISRA C++:2008 Rule 6-6-5, which states:
"'return' statement before end of function 'main'."

Please revise the code below to ensure compliance with Rule 6-6-5:

```
int main() {
   std::vector<std::uint8_t> my_data = {0x01, 0x02, 0x03, 0x04};
   try {
      std::uint8_t simple_checksum = Checksum::simpleChecksum(my_data);
      std::cout << "Simple Checksum: 0x" << std::hex << static_cast<int>(simple_checksum) << std::endl;
      std::uint16_t fletcher_checksum = Checksum::fletcher16(my_data);
      std::cout << "Fletcher-16 Checksum: 0x" << std::hex << fletcher_checksum << std::endl;
      std::uint8_t crc_checksum = Checksum::crc8(my_data);
      std::cout << "CRC-8 Checksum: 0x" << std::hex << static_cast<int>(crc_checksum) << std::endl;
   }
   catch (const std::runtime_error& error) {
      std::cerr << "Error: " << error.what() << std::endl;
      return 1;
   }
   return 0;
}
```

**OUTPUT RESPONSE**

To comply with MISRA C++:2008 Rule 6-6-5, the return 1; statement within the catch block needs to be removed. Instead, a status variable can be used to store the return value, and then main can return this variable at the very end of the function.Here's the revised code:

```
#include <iostream>
#include <vector>
#include <cstdint>
#include <stdexcept> // For std::runtime_error
// Assuming Checksum class and its methods are defined elsewhere
// For demonstration purposes, let's provide a minimal mock-up
namespace Checksum {
   uint8_t simpleChecksum(const std::vector<uint8_t>& data) {
      if (data.empty()) {
         throw std::runtime_error("Input data is empty for simpleChecksum.");
      }
      uint8_t sum = 0;
      for (uint8_t byte : data) {
         sum += byte;
      }
      return sum;
   }
   uint16_t fletcher16(const std::vector<uint8_t>& data) {
      if (data.empty()) {
         throw std::runtime_error("Input data is empty for fletcher16.");
      }
      uint16_t sum1 = 0;
      uint16_t sum2 = 0;
      for (uint8_t byte : data) {
         sum1 = (sum1 + byte) % 255;
         sum2 = (sum2 + sum1) % 255;
      }
      return (sum2 << 8) | sum1;
   }
   uint8_t crc8(const std::vector<uint8_t>& data) {
      if (data.empty()) {
         throw std::runtime_error("Input data is empty for crc8.");
      }
      // This is a very simple (non-standard) CRC8 for demonstration
      uint8_t crc = 0;
      for (uint8_t byte : data) {
         crc ^= byte;
         for (int i = 0; i < 8; ++i) {
            if (crc & 0x80) {
               crc = (crc << 1) ^ 0x07; // Example polynomial 0x07
            } else {
               crc <<= 1;
            }
         }
      }
      return crc;
   }
}
int main() {
   std::vector<std::uint8_t> my_data = {0x01, 0x02, 0x03, 0x04};
   int return_code = 0; // Initialize return_code to 0 (success)
   try {
      std::uint8_t simple_checksum = Checksum::simpleChecksum(my_data);
      std::cout << "Simple Checksum: 0x" << std::hex << static_cast<int>(simple_checksum) << std::endl;
      std::uint16_t fletcher_checksum = Checksum::fletcher16(my_data);
      std::cout << "Fletcher-16 Checksum: 0x" << std::hex << fletcher_checksum << std::endl;
      std::uint8_t crc_checksum = Checksum::crc8(my_data);
      std::cout << "CRC-8 Checksum: 0x" << std::hex << static_cast<int>(crc_checksum) << std::endl;
   }
   catch (const std::runtime_error& error) {
      std::cerr << "Error: " << error.what() << std::endl;
      return_code = 1; // Set return_code to 1 (error)
   }
   return return_code; // Return the stored status at the end of main
}
```

**FIGURE 5.** Input prompt and output response of Gemini to generate the code with MISRA C++:2008 rule 6-6-5 fixed.

**TABLE 7.** Details of MISRA C++:2008 rule violations for the source code generated by Meta AI.

| Rule | Total | Violation |
|---|---|---|
| Rule 5–0–4 | 21 | **Problem:** Implicit conversion from 'short' to 'uint8_t' (aka 'unsigned char') changes signedness of underlying type.<br>**Code:** uint8_t crc = 0xFF; |
| | | **Problem:** Implicit conversion from 'signed char' to 'size_t' (aka 'unsigned int') changes signedness of underlying type.<br>**Code:** for (size_t i = 0; i < length; ++i) { |
| | | **Problem:** Implicit conversion from 'signed char' to 'uint8_t' (aka 'unsigned char') changes signedness of underlying type.<br>**Code:** for (uint8_t j = 0; j < 8; ++j) { |
| | | **Problem:** Implicit conversion from 'signed char' to 'unsigned char' changes signedness of underlying type.<br>**Code:** for (uint8_t j = 0; j < 8; ++j) { |
| | | **Problem:** Implicit conversion from 'unsigned char' to 'short' changes signedness of underlying type.<br>**Code:** if (crc & 0x80) { |
| | | **Problem:** Implicit conversion from 'signed char' to 'unsigned char' changes signedness of underlying type.<br>**Code:** crc = (crc << 1) ^ 0x31; |
| | | **Problem:** Implicit conversion from 'int' to 'uint16_t' (aka 'unsigned short') changes signedness of underlying type.<br>**Code:** uint16_t crc = 0xFFFF; |
| | | **Problem:** Implicit conversion from 'signed char' to 'size_t' (aka 'unsigned int') changes signedness of underlying type.<br>**Code:** for (size_t i = 0; i < length; ++i) { |
| | | **Problem:** Implicit conversion from 'signed char' to 'uint8_t' (aka 'unsigned char') changes signedness of underlying type.<br>**Code:** for (uint8_t j = 0; j < 8; ++j) { |
| | | **Problem:** Implicit conversion from 'signed char' to 'unsigned char' changes signedness of underlying type.<br>**Code:** for (uint8_t j = 0; j < 8; ++j) { |
| | | **Problem:** Implicit conversion from 'unsigned short' to 'int' changes signedness of underlying type.<br>**Code:** if (crc & 0x8000) { |
| | | **Problem:** Implicit conversion from 'short' to 'unsigned short' changes signedness of underlying type.<br>**Code:** crc = (crc << 1) ^ 0x1021; |
| | | **Problem:** Implicit conversion from 'signed char' to 'uint32_t' (aka 'unsigned int') changes signedness of underlying type.<br>**Code:** uint32_t a = 1; |
| | | **Problem:** Implicit conversion from 'signed char' to 'uint32_t' (aka 'unsigned int') changes signedness of underlying type.<br>**Code:** uint32_t b = 0; |
| | | **Problem:** Implicit conversion from 'signed char' to 'size_t' (aka 'unsigned int') changes signedness of underlying type. |
| | | **Code:** for (size_t i = 0; i < length; ++i) { |
| | | **Problem:** Implicit conversion from 'int' to 'unsigned int' changes signedness of underlying type. **Code:** a = (a + data[i]) % 65521; |
| | | **Problem:** Implicit conversion from 'int' to 'unsigned int' changes signedness of underlying type. **Code:** b = (b + a) % 65521; |
| | | **Problem:** Implicit conversion from 'signed char' to 'const uint8_t' (aka 'const unsigned char') changes signedness of underlying type. **Code:** const uint8_t data[] = {0x12, 0x34, 0x56, 0x78}; |
| | | **Problem:** Implicit conversion from 'signed char' to 'const uint8_t' (aka 'const unsigned char') changes signedness of underlying type. **Code:** const uint8_t data[] = {0x12, 0x34, 0x56, 0x78}; |
| | | **Problem:** Implicit conversion from 'signed char' to 'const uint8_t' (aka 'const unsigned char') changes signedness of underlying type. **Code:** const uint8_t data[] = {0x12, 0x34, 0x56, 0x78}; |
| | | **Problem:** Implicit conversion from 'signed char' to 'const uint8_t' (aka 'const unsigned char') changes signedness of underlying type. **Code:** const uint8_t data[] = {0x12, 0x34, 0x56, 0x78}; |
| Rule 5–0–21 | 9 | **Problem:** Bitwise operator '&' applied to signed underlying type. **Code:** if (crc & 0x80) { |
| | | **Problem:** Bitwise operator '^' applied to signed underlying type. **Code:** crc = (crc << 1) ^ 0x31; |
| | | **Problem:** Bitwise operator '<<' applied to signed underlying type. **Code:** crc = (crc << 1) ^ 0x31; |
| | | **Problem:** Bitwise operator '<<=' applied to signed underlying type. **Code:** crc <<= 1; |
| | | **Problem:** Bitwise operator '&' applied to signed underlying type. **Code:** if (crc & 0x8000) { |
| | | **Problem:** Bitwise operator '^' applied to signed underlying type. **Code:** crc = (crc << 1) ^ 0x1021; |
| | | **Problem:** Bitwise operator '<<' applied to signed underlying type. **Code:** crc = (crc << 1) ^ 0x1021; |
| | | **Problem:** Bitwise operator '<<=' applied to signed underlying type. **Code:** crc <<= 1; |
| | | **Problem:** Bitwise operator '<<' applied to signed underlying type. **Code:** return (b << 16) \| a; |
| Rule 7–1–1 | 7 | **Problem:** Parameter 'data' of function 'calculateCrc8 (const uint8_t *, size_t)' could be const. **Code:** uint8_t calculateCrc8 (const uint8_t* data, size_t length) { |
| | | **Problem:** Parameter 'length' of function 'calculateCrc8 (const uint8_t *, size_t)' could be const. **Code:** uint8_t calculateCrc8 (const uint8_t* data, size_t length) { |
| | | **Problem:** Parameter 'data' of function 'calculateCrc16 (const uint8_t *, size_t)' could be const. **Code:** uint16_t calculateCrc16 (const uint8_t* data, size_t length) { |
| | | **Problem:** Parameter 'length' of function 'calculateCrc16(const uint8_t *, size_t)' could be const. **Code:** uint16_t calculateCrc16 (const uint8_t* data, size_t length) { |
| | | **Problem:** Parameter 'data' of function 'calculateAdler32 (const uint8_t *, size_t)' could be const. **Code:** uint32_t calculateAdler32 (const uint8_t* data, size_t length) { |
| | | **Problem:** Parameter 'length' of function 'calculateAdler32 (const uint8_t *, size_t)' could be const. **Code:** uint32_t calculateAdler32 (const uint8_t* data, size_t length) { |
| | | **Problem:** Local variable 'length' could be const. **Code:** size_t length = sizeof(data); |
| Rule 5–0–6 | 4 | **Problem:** Implicit conversion of integer to smaller underlying type ('short' to 'uint8_t' (aka 'unsigned char')). **Code:** uint8_t crc = 0xFF; |
| | | **Problem:** Implicit conversion of integer to smaller underlying type ('short' to 'bool'). **Code:** if (crc & 0x80) { |
| | | **Problem:** Implicit conversion of integer to smaller underlying type ('int' to 'uint16_t' (aka 'unsigned short')). **Code:** uint16_t crc = 0xFFFF; |
| | | **Problem:** Implicit conversion of integer to smaller underlying type ('int' to 'bool'). **Code:** if (crc & 0x8000) { |
| Rule 0–1–3 | 3 | **Problem:** Local variable 'crc8' declared in 'main' not subsequently referenced. **Code:** uint8_t crc8 = calculateCrc8 (data, length); |
| | | **Problem:** Local variable 'crc16' declared in 'main' not subsequently referenced. **Code:** uint16_t crc16 = calculateCrc16 (data, length); |
| | | **Problem:** Local variable 'adler32' declared in 'main' not subsequently referenced. **Code:** uint32_t adler32 = calculateAdler32 (data, length); |
| Rule 0–1–4 | 3 | **Problem:** Local variable 'crc8' declared in 'main' not subsequently referenced. **Code:** uint8_t crc8 = calculateCrc8 (data, length); |
| | | **Problem:** Local variable 'crc16' declared in 'main' not subsequently referenced. **Code:** uint16_t crc16 = calculateCrc16 (data, length); |
| | | **Problem:** Local variable 'adler32' declared in 'main' not subsequently referenced. |
| | | **Code:** uint32_t adler32 = calculateAdler32 (data, length); |
| Rule 0–1–6 | 3 | **Problem:** Last value assigned to 'adler32' not used.<br>**Code:** return 0; |
| | | **Problem:** Last value assigned to 'crc16' not used.<br>**Code:** return 0; |
| | | **Problem:** Last value assigned to 'crc8' not used.<br>**Code:** return 0; |
| Rule 0–1–9 | 3 | **Problem:** Last value assigned to 'adler32' not used.<br>**Code:** return 0; |
| | | **Problem:** Last value assigned to 'crc16' not used.<br>**Code:** return 0; |
| | | **Problem:** Last value assigned to 'crc8' not used.<br>**Code:** return 0; |
| Rule 3–3–1 | 3 | **Problem:** External symbol 'calculateCrc8' could be made static.<br>**Code:** uint8_t calculateCrc8 (const uint8_t* data, size_t length) { |
| | | **Problem:** External symbol 'calculateCrc16' could be made static.<br>**Code:** uint16_t calculateCrc16 (const uint8_t* data, size_t length) { |
| | | **Problem:** External symbol 'calculateAdler32' could be made static.<br>**Code:** uint32_t calculateAdler32 (const uint8_t* data, size_t length) { |
| Rule 5–2–12 | 3 | **Problem:** Array type passed to function expecting a pointer.<br>**Code:** uint8_t crc8 = calculateCrc8 (data, length); |
| | | **Problem:** Array type passed to function expecting a pointer.<br>**Code:** uint16_t crc16 = calculateCrc16 (data, length); |
| | | **Problem:** Array type passed to function expecting a pointer.<br>**Code:** uint32_t adler32 = calculateAdler32 (data, length); |
| Rule 7–3–1 | 3 | **Problem:** Global declaration of symbol 'calculateCrc8'.<br>**Code:** uint8_t calculateCrc8 (const uint8_t* data, size_t length) { |
| | | **Problem:** Global declaration of symbol 'calculateCrc16'.<br>**Code:** uint16_t calculateCrc16 (const uint8_t* data, size_t length) { |
| | | **Problem:** Global declaration of symbol 'calculateAdler32'.<br>**Code:** uint32_t calculateAdler32 (const uint8_t* data, size_t length) { |
| Rule 5–0–10 | 2 | **Problem:** The result of the << operator applied to an object with an underlying type of 'unsigned char' must be cast to 'unsigned char' in this context.<br>**Code:** crc = (crc << 1) ^ 0x31; |
| | | **Problem:** The result of the << operator applied to an object with an underlying type of 'unsigned short' must be cast to 'unsigned short' in this context.<br>**Code:** crc = (crc << 1) ^ 0x1021; |
| Rule 5–0–13 | 2 | **Problem:** Condition of 'if' statement has non-Boolean type 'int'.<br>**Code:** if (crc & 0x80) { |
| | | **Problem:** Condition of 'if' statement has non-Boolean type 'int'.<br>**Code:** if (crc & 0x8000) { |
| Rule 5–0–20 | 1 | **Problem:** Bitwise operator '^=' used with non-constant operands of differing underlying types ('unsigned short' and 'unsigned char').<br>**Code:** crc ^= data[i]; |

Several violations categorized under "Unnecessary Constructs" were identified exclusively in the code generated by Meta AI, and not observed in the outputs from ChatGPT, DeepSeek, or Gemini. These include 3 violations each of Rules 0-1-3, 0-1-4, 0-1-6, and 0-1-9, which indicate the presence of unused variables. Specifically, the violations of Rules 0-1-3 and 0-1-4 relate to the declaration of variables “crc8”, “crc16”, and “adler32”, which are assigned the results of checksum algorithm functions but are never subsequently referenced.

These violations occur within the “main()” function, which serves as test code and is intended to be placed in a separate scope. Similarly, the violations of Rules 0-1-6 and 0-1-9 also occur in “main()” function and result from the final values assigned to “crc8”, “crc16”, and “adler32” not being used i.e., they are neither printed, stored, nor returned.

Meta AI’s code also contains highest number of Rule 5-0-4 violations (21 in total), across all evaluated LLMs. This rule addresses undesirable implicit conversions from signed to unsigned types, which may result in undefined or unintended behavior. Likewise, Meta AI exhibits the greatest number of Rule 5-0-21 violations (9 instances), which pertain to the use of bitwise operators on signed types, an operation that may not yield meaningful results and can compromise software safety and correctness. Additionally, 7 violations of Rule 7-1-1 were found in the “main()” function. This rule mandates that variables which are not intended to be modified must be declared with const qualification.

Meta AI's output also includes 3 violations each of the Rules 3-3-1 and 5-2-12, which were not present in code generated by the other LLMs. Rule 3-3-1 relates to the scope and linkage of objects and functions, recommending that symbols be declared in a header file when external linkage is needed, or otherwise within an unnamed namespace or marked static to limit their visibility. This practice enhances encapsulation and minimizes the risk of symbol collisions. Rule 5-2-12 violations occur in the “main()” function, where the functions “calculateCrc8 (data, length)”, “calculateCrc16 (data, length)”, and “calculateAdler32 (data, length)” receive “data[]” as an argument, despite expecting a pointer. When an array decays to a pointer, its bounds are lost, which can lead to safety and correctness issues in critical software.

Additional violations include 4 instances of Rule 5-0-6, which prohibits implicit conversion from a larger integer type to a smaller one due to the potential for data loss. Furthermore, 3 violations of Rule 7-3-1 were identified, arising from global declaration of "calculateCrc8", "calculateCrc16", and "calculateAdler32". This rule discourages placing user-defined identifiers in global namespace, except for "main()" or specific "extern C" declarations, due to the risk of name collisions, reduced modularity, and diminished maintainability, especially in large-scale or safety-critical systems. MISRA C++:2008 recommends encapsulating such entities within namespaces or classes to improve code organization and clarity.

Two violations of Rule 5-0-13 were identified, which stipulates that condition of an "if" statement must be of "Boolean" type; in both instances, the condition is of type "int", thereby violating the rule. Additionally, one violation of Rule 5-0-20 was observed, where bitwise operator "^=" is applied to non-constant operands with differing underlying types, specifically "unsigned short" and "unsigned char", which can lead to type mismatch issues and unpredictable behavior.

**INPUT PROMPT**

The following line of code:
return 0;

violates MISRA C++:2008 Rule 0-1-6, which states:
"Last value assigned to 'crc8' not used."

Please revise the code below to ensure compliance with Rule 0-1-6:

```
int main() {
    const uint8_t data[] = {0x12, 0x34, 0x56, 0x78};
    size_t length = sizeof(data);
    uint8_t crc8 = calculateCrc8(data, length);
    uint16_t crc16 = calculateCrc16(data, length);
    uint32_t adler32 = calculateAdler32(data, length);
    // Print or verify the checksum values
    return 0;
}
```

**OUTPUT RESPONSE**

To comply with MISRA C++:2008 Rule 0-1-6, you should use the last assigned value to the variable 'crc8'. Here's the revised code:

```
int main() {
    const uint8_t data[] = {0x12, 0x34, 0x56, 0x78};
    size_t length = sizeof(data);
    uint8_t crc8 = calculateCrc8(data, length);
    uint16_t crc16 = calculateCrc16(data, length);
    uint32_t adler32 = calculateAdler32(data, length);
    // Use the last assigned value to crc8
    return crc8;
}
```

In this revised code, the last assigned value to 'crc8' is used as the return value of the main function, thus ensuring compliance with MISRA C++:2008 Rule 0-1-6.

**FIGURE 6. Input prompt and output response of Meta AI to generate the code with MISRA C++:2008 rule 0-1-6 fixed.**

Finally, 2 violations of Rule 5-0-10 were reported, which requires that the result of applying "<<" operator to objects with underlying types such as "unsigned char" or "unsigned short" must be explicitly cast back to the original type. This ensures type safety and prevents unintended side effects in bitwise operations.

To further evaluate Meta AI's capability to address MISRA C++:2008 non-compliance, a targeted prompt was issued specifically to rectify the Rule 0-1-6 violation. In response, Meta AI generated a revised version of the code that successfully resolved the identified instance of non-compliance, as shown in FIGURE 6. However, it failed to identify and correct two additional similar violations involving the "crc16" and "adler32" variables. Moreover, the corrected output did not ensure full adherence to other applicable MISRA C++:2008 rules. For instance, it did not include explicit casting of the return value to int, potentially resulting in a violation of type conversion requirements. This highlights the need for more comprehensive verification mechanisms when using LLMs in safety-critical software.

### *F. ANALYSIS OF MICROSOFT COPILOT*

The source code generated by Microsoft Copilot comprises 68 lines and exhibits 25 instances of non-compliance with 9 distinct rules of the MISRA C++:2008 standard. Specifically, the output includes 7 violations of Rule 7-1-1, 3 violations each of Rules 3-3-1 and 7-3-1, and 2 violations each of Rules 0-1-3, 0-1-4, 5-0-3, 5-0-10, 5-0-20, and 5-2-12. Notably, Copilot does not generate any violations of Rule 5-0-4, which concerns implicit conversions from signed to unsigned types. This contrasts with the outputs of all other evaluated LLMs, which exhibited multiple violations of this rule.

A detailed summary of the identified violations is provided in TABLE 8. with the corresponding sections of code underlined and highlighted in red to clearly illustrate the specific instances of non-compliance.

**TABLE 8. Details of MISRA C++:2008 rule violations for the source code generated by Microsoft Copilot.**

| Rule | Total | Violation |
|---|---|---|
| Rule 7–1–1 | 7 | **Problem:** Parameter 'data' of function 'calculateXORChecksum (const std::uint8_t *, std::size_t)' could be const.<br>**Code:** std::uint8_t calculateXORChecksum (const std::uint8_t* data, std::size_t length) |
| | | **Problem:** Parameter 'length' of function 'calculateXORChecksum (const std::uint8_t *, std::size_t)' could be const.<br>**Code:** std::uint8_t calculateXORChecksum (const std::uint8_t* data, std::size_t length) |
| | | **Problem:** Parameter 'value' of function 'reflectBits (std::uint32_t, std::uint8_t)' could be const.<br>**Code:** std::uint32_t reflectBits(std::uint32_t value, std::uint8_t bitCount) |
| | | **Problem:** Parameter 'bitCount' of function 'reflectBits (std::uint32_t, std::uint8_t)' could be const.<br>**Code:** std::uint32_t reflectBits(std::uint32_t value, std::uint8_t bitCount) |
| | | **Problem:** Parameter 'data' of function 'calculateCRC32 (const std::uint8_t *, |
| | | std::size_t)' could be const.<br>**Code:** std::uint32_t calculateCRC32(const std::uint8_t* data, std::size_t length) |
| | | **Problem:** Parameter 'length' of function 'calculateCRC32(const std::uint8_t *, std::size_t)' could be const.<br>**Code:** std::uint32_t calculateCRC32(const std::uint8_t* data, std::size_t length) |
| | | **Problem:** Local variable 'byte' could be const.<br>**Code:** std::uint32_t byte = reflectBits (data[i], 8U); |
| Rule 3–3–1 | 3 | **Problem:** External symbol 'reflectBits' could be made static.<br>**Code:** std::uint32_t reflectBits (std::uint32_t value, std::uint8_t bitCount) |
| | | **Problem:** External symbol 'calculateCRC32' could be made static.<br>**Code:** std::uint32_t calculateCRC32 (const std::uint8_t* data, std::size_t length) |
| | | **Problem:** External symbol 'calculateXORChecksum' could be made static.<br>**Code:** std::uint8_t calculateXORChecksum (const std::uint8_t* data, std::size_t length) |
| Rule 7–3–1 | 3 | **Problem:** Global declaration of symbol 'calculateXORChecksum'.<br>**Code:** std::uint8_t calculateXORChecksum (const std::uint8_t* data, std::size_t length) |
| | | **Problem:** Global declaration of symbol 'reflectBits'.<br>**Code:** std::uint32_t reflectBits (std::uint32_t value, std::uint8_t bitCount) |
| | | **Problem:** Global declaration of symbol 'calculateCRC32'.<br>**Code:** std::uint32_t calculateCRC32 (const std::uint8_t* data, std::size_t length) |
| Rule 0–1–3 | 2 | **Problem:** Local variable 'xorChecksum' declared in 'main' not subsequently referenced.<br>**Code:** const std::uint8_t xorChecksum = calculateXORChecksum (data, length); |
| | | **Problem:** Local variable 'crc32' declared in 'main' not subsequently referenced.<br>**Code:** const std::uint32_t crc32 = calculateCRC32 (data, length); |
| Rule 0–1–4 | 2 | **Problem:** Local variable 'xorChecksum' declared in 'main' not subsequently referenced.<br>**Code:** const std::uint8_t xorChecksum = calculateXORChecksum (data, length); |
| | | **Problem:** Local variable 'crc32' declared in 'main' not subsequently referenced.<br>**Code:** const std::uint32_t crc32 = calculateCRC32 (data, length); |
| Rule 5–0–3 | 2 | **Problem:** Implicit conversion of underlying type of integer cvalue expression from 'unsigned char' to 'unsigned int'.<br>**Code:** if ((value & (1U << i)) != 0U) |
| | | **Problem:** Implicit conversion of underlying type of integer cvalue expression from 'unsigned char' to 'unsigned int'.<br>**Code:** reflection \|= (1U << (bitCount - 1U - i)); |
| Rule 5–0–10 | 2 | **Problem:** The result of the << operator applied to an object with an underlying type of 'unsigned char' must be cast to 'unsigned char' in this context.<br>**Code:** if ((value & (1U << i)) != 0U) |
| | | **Problem:** The result of the << operator applied to an object with an underlying type of 'unsigned char' must be cast to 'unsigned char' in this context<br>**Code:** reflection \|= (1U << (bitCount - 1U - i)); |
| Rule 5–0–20 | 2 | **Problem:** Bitwise operator '&' used with non-constant operands of differing underlying types ('unsigned int' and 'unsigned char').<br>**Code:** if ((value & (1U << i)) != 0U) |
| | | **Problem:** Bitwise operator '\|=' used with non-constant operands of differing underlying types ('unsigned int' and 'unsigned char').<br>**Code:** reflection \|= (1U << (bitCount - 1U - i)); |
| Rule 5–2–12 | 2 | **Problem:** Array type passed to function expecting a pointer.<br>**Code:** const std::uint8_t xorChecksum = calculateXORChecksum (data, length); |
| | | **Problem:** Array type passed to function expecting a pointer.<br>**Code:** const std::uint32_t crc32 = calculateCRC32 (data, length); |

The code generated by Microsoft Copilot includes 7 violations of Rule 7-1-1, which mandates that any variable or parameter not intended to be modified should be declared with const qualification to prevent unintended changes. For instance, while the pointer "data" is correctly declared to point to "const std::uint8_t", thereby protecting the underlying data from modification, the pointer itself is not declared as "const". As a result, it remains mutable and could be reassigned within the function scope, violating the intent of the rule.

Additionally, 3 violations of Rule 3-3-1 were observed in the code generated by Copilot. This rule concerns declarative regions and scope, requiring that objects or functions be declared in a header file if external linkage is intended. Otherwise, such entities should be declared within an unnamed namespace or marked as static to limit their visibility. This practice enhances encapsulation, reduces symbol conflicts, and improves maintainability in safety-critical software.

Furthermore, there are 3 violations of Rule 7-3-1, arising from the global declarations of the functions "reflectBits", "calculateCRC32", and "calculateXORChecksum". Rule 7-3-1 discourages placing user-defined identifiers, such as functions and variables, directly in global namespace, with the exception of "main()" and specific "extern C" cases. Excessive use of global namespace increases the risk of naming collisions, reduces modularity, and complicates maintenance, especially in large-scale or safety-critical projects. MISRA C++:2008 encourages encapsulating such entities within namespaces or classes to promote better structure and avoid ambiguity.

Moreover, 4 violations related to "Unnecessary Constructs" were identified in the code generated by Copilot, comprising 2 violations each of Rules 0-1-3 and 0-1-4. These violations point to the presence of unused variables or dead code, which were not observed in the outputs of ChatGPT, DeepSeek, or Gemini. Specifically, the variables "xorChecksum" and "crc32" are assigned the results of checksum function calls but are never subsequently used. These instances occur within the "main()" function, which serves as test or reference code and is expected to be implemented in an isolated scope.

Two violations of Rule 5-0-3 were identified exclusively in the code generated by ChatGPT and Microsoft Copilot. This rule mandates that a "cvalue" expression must not be implicitly converted to a different type, ensuring type consistency throughout an expression to prevent unintended behavior.

Two violations of Rule 5-0-10 were also observed. This rule requires that the result of the "<<" operator, when applied to an object with an underlying type of "unsigned short" or "unsigned char", must be explicitly cast back to the operand's underlying type. Failure to do so may result in incorrect or unsafe bitwise operations.

**INPUT PROMPT**

The following line of code:
```
const std::uint8_t xorChecksum = calculateXORChecksum(data, length);
```

violates MISRA C++:2008 Rule 0-1-3, which states:
"local variable 'xorChecksum' declared in 'main' not subsequently referenced."

Please revise the code below to ensure compliance with Rule 0-1-3:
```
int main(){
   const std::uint8_t data[] = {0x12U, 0x34U, 0x56U, 0x78U};
   const std::size_t length = sizeof(data) / sizeof(data[0]);
   const std::uint8_t xorChecksum = calculateXORChecksum(data, length);
   const std::uint32_t crc32 = calculateCRC32(data, length);
   // Example usage output
   // The outputs should be sent to an appropriate interface in an embedded system.
   return 0;
}
```

**OUTPUT RESPONSE**

To comply with MISRA C++:2008 Rule 0-1-3, each local variable must be used after its declaration. In your code, xorChecksum is calculated but not used or referenced later, which violates the rule.

You can fix this by making some form of use of xorChecksum—for example, by assigning it to a dummy variable, sending it to an output interface, or using it in a computation. For illustration purposes, here's a revised version that outputs the value using a placeholder comment (assuming this will be replaced with actual embedded system interface logic):

```
#include <cstddef>
#include <cstdint>

std::uint8_t calculateXORChecksum(const std::uint8_t* data, std::size_t length);
std::uint32_t calculateCRC32(const std::uint8_t* data, std::size_t length);

int main() {
   const std::uint8_t data[] = {0x12U, 0x34U, 0x56U, 0x78U};
   const std::size_t length = sizeof(data) / sizeof(data[0]);
   const std::uint8_t xorChecksum = calculateXORChecksum(data, length);
   const std::uint32_t crc32 = calculateCRC32(data, length);

   // Example usage - Replace with actual embedded system interface output
   (void)xorChecksum; // Prevents unused variable warning satisfying Rule 0-1-3
   (void)crc32;

   return 0;
}
```

The cast to (void) is a common way in C/C++ to mark a variable as "used" when its value doesn't need to be accessed explicitly. It satisfies MISRA's requirement for a reference while signaling intentional suppression of use.

**FIGURE 7. Input prompt and output response of Microsoft Copilot to generate code with MISRA C++:2008 rule 0-1-3 fixed.**

Additionally, 2 violations of Rule 5-0-20 were reported, wherein the bitwise operators "&" and "|=" were used with non-constant operands of differing underlying types, specifically, "unsigned int" and "unsigned char". Such usage may lead to type mismatches and undefined behavior.

The code generated by Copilot exhibited 2 violations of Rule 5-2-12, which were not observed in the outputs from ChatGPT, DeepSeek, or Gemini. These violations occur within the "main()" function and involve the functions "calculateCRC32 (data, length)" and "calculateXORChecksum (data, length)" being called with an argument of array type "data[]", whereas the functions expect a pointer. When an array decays to a pointer, its bounds are lost, potentially introducing risks in memory access and undermining code safety and predictability.

To further evaluate Microsoft Copilot's ability to address non-compliance with MISRA C++:2008, a focused prompt was issued targeting the correction of Rule 0-1-3 violation. In response, Copilot successfully produced a corrected version of the code and generated a complete implementation that resolved the non-compliance, as illustrated in FIGURE 7. Notably, it also correctly identified and addressed a similar violation involving the "crc32" variable, demonstrating its capacity to generalize corrections beyond a single instance.

### *G. ANALYSIS OF MISRA C++:2008 VIOLATION FIXING*

An additional experiment was conducted to evaluate the ability of LLMs to identify and rectify specific MISRA C++:2008 rule violations within generated code. The LLMs evaluated in this study included OpenAI ChatGPT (GPT-4o mini), DeepSeek-V3, Google Gemini 2.5 Flash, Meta AI (LLaMA 4), and Microsoft Copilot. The focus was placed on Rules 5-0-4, 7-1-1, and 5-0-10, as these rules were commonly violated across all models.

Each LLM was provided with a standardized prompt in the following format: "This C++ code contains multiple violations of MISRA C++:2008 Rule [RULE_NO], which states that '[RULE_DESCRIPTION]'. Please identify and fix all those violations and output the complete fixed code with detailed explanation of fixes.". The outcomes of this experiment are presented in TABLE 9. TABLE 10. and TABLE 11. corresponding to the Rules 5-0-4, 7-1-1, and 5-0-10, respectively.

The results indicate that ChatGPT successfully identified, corrected, and explained all violations of the Rules 5-0-4 and 5-0-10. There were no instances of Rule 7-1-1 violations in the code generated by ChatGPT, as confirmed by PC-lint Plus SAT. DeepSeek failed to identify or correct violations of Rule 5-0-4, but successfully resolved all instances of Rules 7-1-1 and 5-0-10. Google Gemini was able to identify, correct, and explain approximately half of the violations for Rules 5-0-4 and 5-0-10, while demonstrating complete accuracy for Rule 7-1-1 in its own generated code.

Meta AI successfully addressed all the violations of the Rule 5-0-10, identified 4 out of 21 violations for Rule 5-0-4, and failed to correctly fix or explain 3 out of 7 violations of Rule 7-1-1, despite correctly identifying them. Microsoft

Copilot correctly identified, fixed, and explained all violations of Rule 5-0-10, and resolved 1 out of 7 violations of Rule 7-1-1. Notably, no violations of Rule 5-0-4 were present in the code generated by Copilot, as verified by PC-lint Plus SAT.

**TABLE 9.** Evaluation of LLMs' effectiveness in addressing MISRA C++:2008 rule 5-0-4 violations

| LLM | Explained | Identified | Fixed |
|---|---|---|---|
| ChatGPT | Yes | 8/8 | 8/8 |
| DeepSeek | No | 0/5 | 0/5 |
| Gemini | Partial | 5/10 | 5/10 |
| Meta AI | Partial | 4/21 | 4/21 |
| Copilot | N/A | N/A | N/A |

**TABLE 10.** Evaluation of LLMs' effectiveness in addressing MISRA C++:2008 rule 7-1-1 violations

| LLM | Explained | Identified | Fixed |
|---|---|---|---|
| ChatGPT | N/A | N/A | N/A |
| DeepSeek | Yes | 6/6 | 6/6 |
| Gemini | Yes | 7/7 | 7/7 |
| Meta AI | Partial | 7/7 | 4/7 |
| Copilot | Partial | 1/7 | 1/7 |

**TABLE 11.** Evaluation of LLMs' effectiveness in addressing MISRA C++:2008 rule 5-0-10 violations

| LLM | Explained | Identified | Fixed |
|---|---|---|---|
| ChatGPT | Yes | 1/1 | 1/1 |
| DeepSeek | Yes | 1/1 | 1/1 |
| Gemini | Partial | 1/2 | 1/2 |
| Meta AI | Yes | 2/2 | 2/2 |
| Copilot | Yes | 2/2 | 2/2 |

## IV. RELATED WORK

Hansson and Ellréus [12] conducted an empirical study evaluating the correctness and quality of code generated by ChatGPT and GitHub Copilot. Their work highlights both the potential and the limitations of AI-assisted coding, emphasizing challenges related to reliability, consistency, and the need for human oversight. Chen et al. [13] presented a comprehensive evaluation of large language models trained on code, examining their ability to solve programming problems across diverse benchmarks. The study demonstrated strong performance in code generation tasks while also identifying limitations in reasoning, correctness, and generalization. Yetistiren et al. [14] assessed the quality of code generated by GitHub Copilot, focusing on aspects such as correctness, readability, and maintainability. Their findings indicated that while Copilot can assist developers by producing functional code, it often requires human oversight to address issues of accuracy and quality.

Wang and Chen [15] provided a review of code generation using large language models (LLMs), examining their applications, benefits, and evaluation methods. The study highlighted both the potential of LLMs in improving software development efficiency and the challenges related to reliability, correctness, and evaluation frameworks. Li et al. [16] investigated fine-tuning large language models to enhance secure code generation, demonstrating that tailored tuning approaches can reduce vulnerabilities and improve the reliability of generated code. Their work highlights the potential of adapting foundation models for security-aware software engineering tasks. He et al. [17] explored instruction tuning as a method to improve secure code generation, showing that aligning LLMs with security-focused instructions reduces vulnerabilities in the produced code. Their study demonstrates the effectiveness of instruction tuning for enhancing standards- and security-aware code synthesis.

Jiang et al. [18] provide a comprehensive literature survey on LLMs for code generation, introducing a taxonomy of recent developments, from data sourcing to ethical implications, and empirically comparing model performance using benchmarks like HumanEval and MBPP. Dai et al. [19] present a comprehensive evaluation of four secure code generation techniques, assessing security and functionality simultaneously on unified code samples using multiple static analyzers and two large language models. Their findings reveal that while these methods can reduce detected vulnerabilities, they often degrade functionality, underscoring the need for more rigorous frameworks.

## V. LIMITATIONS AND FUTURE WORK

In this study, publicly available Large Language Models (LLMs) were utilized due to the substantial computational and resource requirements associated with training such models from scratch. Consequently, these models were treated as closed-box systems, limiting the ability to perform in-depth analysis or interpret certain results with full transparency. Moreover, as I had no direct control over model updates or versioning, it is possible that future studies may yield different outcomes should the underlying models be modified.

A limitation of this study is that the analysis was conducted using MISRA C++:2008 rather than the latest MISRA C++:2023 standard. The primary reason for this choice was the restricted availability and commercial licensing of the 2023 specification, which is not freely accessible for academic research. Furthermore, the Static Analysis Tool (SAT) employed in this study—PC-lint Plus (version 2.2)—currently supports compliance checking only against MISRA C++:2008, and does not yet include rule definitions or configuration files for MISRA C++:2023. Consequently, reproducing the analysis using the latest version was not technically feasible within the current toolchain. Despite this limitation, MISRA C++:2008 remains widely recognized and actively used in safety-critical industries, particularly for legacy and certified systems developed under standards such as DO-178C and ISO 26262. While MISRA C++:2023 introduces updates aligned with newer C++ features (e.g., auto, constexpr, and smart pointers) and enhanced alignment with CERT C++ and AUTOSAR C++14, its core safety intent and compliance philosophy remain consistent with the 2008 edition. Therefore, the findings and conclusions of this study remain valid, with future work planned to extend the evaluation to MISRA C++:2023 once both the standard and supporting analysis tools become available for academic use.

This investigation focused on five widely used LLMs, the C++ programming language, MISRA C++:2008 standard, and employed PC-lint Plus Static Analysis Tool (SAT) along with a representative example implementing common checksum algorithms. Future studies may expand on this foundation by incorporating a broader range of LLMs, evaluating additional code examples, applying multiple SATs, exploring alternative programming languages, or comparing adherence to different industry coding standards.

Further research is encouraged to investigate the performance of domain-specific or code-optimized LLMs, such as GitHub Copilot, and contrast their outputs with those of general-purpose models evaluated in this study. Additionally, formulation of input prompts remains inherently subjective and can vary significantly between researchers. Exploring the impact of prompt engineering on code quality and compliance remains a promising direction for future work.

Finally, this study is based on a set of assumptions and reflects a typical usage pattern whereby developers employ LLMs to generate functional code. Future investigations should consider the diverse strategies adopted by developers of varying expertise, goals, and workflows when using LLMs for software development tasks.

## VI. CONCLUSION

Large Language Models (LLMs) are increasingly being adopted to automate various aspects of software development, with code generation emerging as one of their most impactful applications. The integration of LLMs into software engineering workflows holds transformative potential, enhancing coding efficiency, accelerating prototyping, and fostering innovation in both individual and enterprise-level development environments. However, the utility of LLMs in safety-critical domains necessitates careful scrutiny. In such contexts, adherence to software development standards such as MISRA C++, mandated by certification frameworks like DO-178C for avionics systems is critical. This raises a fundamental question: Can LLMs reliably generate MISRA C++ compliant code suitable for use in safety-critical applications?

To investigate this question, this study evaluates the degree of compliance of LLM-generated C++ code with the MISRA C++:2008 standard. While some commercially available LLMs are designed specifically for code generation, the focus of this work is on publicly accessible and general-purpose models. The five LLMs selected for this analysis include OpenAI ChatGPT (GPT-4o mini), DeepSeek-V3, Google Gemini 2.5 Flash, Meta AI (LLaMA 4), and Microsoft Copilot. Each model was tasked with generating C++ code to implement commonly used checksum algorithms, based on prompts explicitly requesting MISRA-compliant output.

The generated code was compiled using Eclipse IDE. It was observed that the code generated by DeepSeek failed to compile due to two programming errors, whereas the outputs from Meta AI and Microsoft Copilot compiled successfully but produced three and two compiler warnings, respectively, primarily related to unused variables. To assess compliance, the source code was analyzed using the industry-recognized static analysis tool PC-lint Plus, configured for MISRA C++:2008 rule checking. The findings revealed that none of the evaluated models produced code that was fully compliant with the MISRA standard, despite clear and specific prompts. DeepSeek exhibited the fewest violations (13), while Meta AI generated the most violations (67).

Subsequently, all models were given a focused prompt instructing them to correct a specific MISRA rule violation within a given code snippet. Each LLM successfully addressed the specified violation and returned corrected code along with explanations. In a further step, the models were asked to identify and fix all violations of selected MISRA rules within the complete generated code. ChatGPT demonstrated the highest consistency, successfully resolving all targeted rule violations. The other LLMs showed partial success, with mixed results in identifying and rectifying violations across different rule sets.

These findings suggest that, while LLMs are capable of generating syntactically valid and functionally structured code, they do not reliably ensure MISRA C++ compliance. This limitation stems from both architectural and methodological constraints inherent to current LLMs. Compliance with MISRA C++ requires enforcement of strict syntactic and semantic rules, including type safety, scope management, control and data flow analysis, and inter-file consistency. These capabilities are beyond the probabilistic, token-based nature of LLMs. Since LLMs lack built-in mechanisms for static or semantic analysis, they are unable to guarantee adherence to rules that require deep contextual awareness or cross-referencing of definitions and declarations.

Furthermore, LLMs are typically trained on open-source codebases that are not authored according to MISRA standards, thereby inheriting and reproducing common programming practices that often conflict with MISRA guidelines such as the use of implicit type conversions, dynamic memory allocation, unrestricted inheritance, or reliance on global state. Even when explicitly instructed to follow MISRA rules, LLMs may overlook nuanced or less apparent rule violations due to the absence of internal compliance validation mechanisms.

In conclusion, while LLMs can serve as effective tools for initial code generation and prototyping, they cannot currently be relied upon to produce fully MISRA C++-compliant code for safety-critical systems. Their use in such domains must therefore be complemented by rigorous static analysis and manual review processes to ensure compliance with industry standards and regulatory requirements.

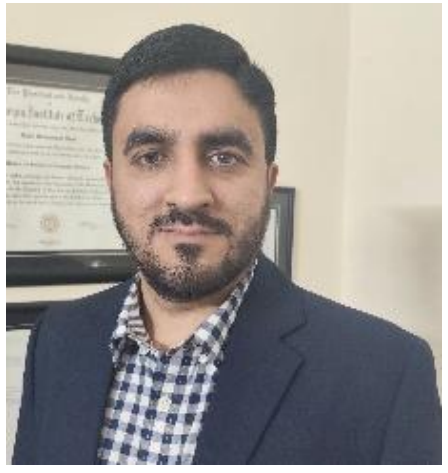

**MALIK MUHAMMAD UMER** received his bachelor's degree in Aeronautical Engineering (Avionics) from National University of Sciences and Technology, Pakistan and the master's degree in Computer Science from Georgia Institute of Technology, USA. Currently, he is an Independent Volunteer Researcher affiliated with the Georgia Institute of Technology in an online capacity, and serves as the Deputy Director of the Software Development Group within the Aerospace Industry, Pakistan. His research interests include software development, particularly software testing.